\documentclass[apj]{emulateapj}
\usepackage{dsfont}
\usepackage{multirow}

\newcommand{\myemail}{havenhaus@astro.phys.ethz.ch}

\slugcomment{Accepted for publication in ApJ on November 25, 2013}

\shorttitle{Structures in the protoplanetary disk of HD142527 seen in polarized scattered light}
\shortauthors{Avenhaus et al.}

\begin{document}

\title{Structures in the protoplanetary disk of HD142527 seen in polarized scattered light}

\author{Henning Avenhaus$^{1,2}$, Sascha P. Quanz$^2$, Hans Martin Schmid$^2$, Michael R. Meyer$^2$, Antonio Garufi$^2$, Sebastian Wolf$^3$, and Carsten Dominik$^4$}
\email{\myemail}

\altaffiltext{1}{Based on observations collected at the European Organisation for Astronomical Research in the Southern Hemisphere, Chile, under program number 089.C-0611(A).}
\altaffiltext{2}{ETH Zurich, Institute for Astronomy, Wolfgang-Pauli-Strasse 27, 8093 Zurich, Switzerland}
\altaffiltext{3}{University of Kiel, Institute of Theoretical Physics and Astrophysics, Leibnizstrasse 15, 24098 Kiel, Germany}    
\altaffiltext{4}{University of Amsterdam, Astronomical Institute Anton Pannekoek, Postbus 94249, 1090 GE Amsterdam, Netherlands}

\begin{abstract}
We present $H$- and $K_{\rm s}$-band polarized differential images (PDI) of the Herbig Ae/Be star HD142527, revealing its optically thick outer disk and the nearly empty gap. The very small inner working angle ($\sim$0.1$\arcsec$) and high resolution achievable with an 8m-class telescope, together with a careful polarimetric calibration strategy, allow us to achieve images that surpass the quality of previous scattered light images. Previously known substructures are resolved more clearly and new structures are seen. Specifically, we are able to resolve 1) half a dozen spiral structures in the disk, including previously known outer-disk spirals as well as new spiral arms and arcs close to the inner rim of the disk; 2) peculiar holes in the polarized surface brightness at position angles of $\sim$0$^\circ$ and $\sim$160$^\circ$; 3) the inner rim on the eastern side of the disk; 4) the gap between the outer and inner disk, ranging from the inner working angle of 0.1$\arcsec$ out to between 0.7 and 1.0$\arcsec$, which is nearly devoid of dust. We then use a Markov-chain Monte-Carlo algorithm to determine several structural parameters of the disk, using very simple assumptions, including its inclination, eccentricity, and the scale height of the inner rim. We compare our results to previous work on this object, and try to produce a consistent picture of the system and its transition disk.
\end{abstract}

\keywords{stars: pre-main sequence --- stars: formation --- protoplanetary disks --- planet-disk interactions --- stars: individual (HD142527)}
\objectname{HD142527} 

\section{Introduction}

Understanding the initial conditions and dominant physical processes of planet formation is vital to explaining the diversity of planetary system architectures observed. This requires the study of circumstellar disks.

The search for and investigation of young stars and their protoplanetary disks has a long history and has been carried out in a variety of different wavelengths. Besides ultraviolet studies, which can be used to measure the accretion rate of the central star, two important regimes can be distinguished: The thermal emission of the disk, and the reflected light. Thermal emission can be used to determine the dominant constituents of a disk through spectroscopy, and it can be used to image the disk, with the short wavelengths tracing the hot, inner material of the disk and the long wavelengths tracing the colder material further out. At short wavelengths ($\lesssim$ 5$\mu$m) and large separations ($>$ a few AU), the light from the disk is dominated by the stellar light scattered off the dust grains. This does not depend on the temperature of the grains, but on their scattering properties. Last but not least, molecular line tracers, mostly in the (sub-)mm, can give us information about the gas rather than the dust in the disk, so we have a rich set of tools to study these circumstellar, planet-forming environments.

Unfortunately, the Herbig Ae/Be and T-Tauri stars harboring them are typically located at 100pc or more from the sun, which means that the physical separations where most planets form ($\lesssim$ 50 AU) translate to angular separations of 0.5$\arcsec$ or less. Probing this regime at high resolution is difficult both in scattered light and thermal emission. In scattered light, PSF (point spread function) subtraction and coronagraphs are often used to to block the stellar light, but while this is an excellent tool for studying the outer parts of such disks, the inner working angle is typically not good enough to probe the circumstellar environment at separations smaller than 0.5-1$\arcsec$ \citep[e.g.][]{grady2001, fukagawa2010, casassus2012}. On the other hand, the ratio of stellar to disk flux is much more favorable in the far-infrared and sub-mm regime, but the telescope resolution criterion usually severely limits the achievable resolution, though ALMA \citep{wootten2009} brings dramatic improvements here.

Polarimetric differential imaging (PDI) is a powerful technique to suppress the stellar light in scattered light observations and has recently been successfully applied to several protoplanetary disks \citep[e.g.,][]{quanz2011b, hashimoto2011, muto2012, quanz2012, hashimoto2012, tanii2012, kusakabe2012, mayama2012, grady2013, quanz2013, garufi2013}. It uses the fact that while the light from the central source is largely unpolarized, scattering on the dust grains in the disk produces polarization. PDI allows to probe the circumstellar environment of young stars very close to the star and achieve the high contrast ratios required to detect their circumstellar disks.

In this paper, we present $H$- and $K_{\rm s}$-band PDI observations of HD142527 with an inner working angle of less than 0.1$\arcsec$ ($\sim$ 15 AU), revealing the large gap in scattered light in unprecedented clarity and showing the outer disk in unprecedented detail, including several previously unrecognized spiral arms and gaps. We discuss our results, present estimations for the basic parameters of the disk using an Markov-chain Monte-Carlo code and link our findings to the current knowledge about this extraordinary disk.

The paper is structured as follows: In section 2, we describe the HD142527 system in general. In Section 3, we give a short introduction to the observations and data reduction. We present the results in Section 4, followed by an analysis of these results in Section 5. In Section 6, we discuss our findings and compare them to earlier studies. We draw our conclusions in Section 7. A detailed description of our data reduction pipeline, which has also been used for the data reduction of our observations of HD169142 and SAO206462 \citep[][]{quanz2013, garufi2013}, is given in the appendix.

\begin{deluxetable}{lcc}
\centering
\tablecaption{Basic parameters of HD142527. 
\label{table:HD142527}}           %
\tablehead{
\colhead{Parameter} & \colhead{Value for HD142527} & \colhead{Reference\tablenotemark{a}}
}
\startdata
RA (J2000) & 15$^h$56$^m$41$^s$.89  & (1) \\ 
DEC (J2000) & -42$^\circ$19$'$23$''$.27   & (1)\\
$J$ [mag] & 6.50$\pm 0.03$  & (1)\\
$H$ [mag]& 5.72$\pm 0.03$   &(1)\\
$K_{\rm s}$ [mag]& 4.98$\pm 0.02$  & (1)\\
$A_V$ [mag]& 0.60$\pm$0.05 / 1.49 & (4),(5)\\
Sp. Type & F6IIIe / F7IIIe & (2),(6) \\
Age [Myr]& $2^{+2}_{-1}$ / $5^{+8}_{-3}$ / $1.0^{+1.5}_{-0.6}$ & (3),(4),(6) \\
$T_{\rm eff}$ [K] & 6300 / 6250 & (2),(4)\\
Mass [M$_\sun$] & 3.5 / 1.9$\pm$0.3 / 2.2$\pm$0.3 & (2),(3),(4)\\ 
Disk mass [M$_\sun$] & 0.1 / 0.15 & (4),(9)\\
Disk inclination & 20$\pm$2$^\circ$ & (10)\\
$R_*$  [R$_\sun$]& 3.8$\pm$0.3 & (4) \\
$L_*$ [L$_\sun$]& $20^{+2}_{-2}$ / $69^{+48}_{-24}$ / 29  & (4),(5),(6) \\
$\dot{M}$ [${\rm M}_\sun{\rm yr}^{-1}$]& 6.9$\cdot10^{-8}$  & (7) \\
Distance [pc] & $230^{+70}_{-40}$ / 145\tablenotemark{b} & (8),(11/12)\\

\enddata
\tablenotetext{a}{References --- (1) 2MASS point source catalog \citep{cutri2003}, (2) \citet{waelkens1996}, (3) \citet{fukagawa2006}, (4) \citet{verhoeff2011}, (5) \citet{vandenancker1998}, (6) \citet{vanboekel2005}, (7) \citet{lopez2006}, (8) \citet{leeuwen2007}, (9) \citet{acke2004a}, (10) \citet{pontoppidan2011}, (11) \citet{acke2004b}, (12) \citet{zeeuw1999}.}
\tablenotetext{b}{adopted value in this paper}
\end{deluxetable}

\section{The HD142527 system}

HD142527 is an F-type Herbig Ae/Be star \citep{waelkens1996, malfait1998} that has been classified as group Ia \citep{meeus2001}, meaning that its spectral energy disribution (SED) rises in the mid-IR and also shows a 10$\mu$m silicate emission feature. It further shows signatures of water ice \citep{malfait1999, honda2009}. The circumstellar environment consists of an inner and outer disk separated by a gap which has been detected in the mid-IR and from sub-mm continuum observations. While the inner border of the outer disk is well-constrained at $\sim$ 130 AU and is eccentric, the extent of the inner disk is observationally less constrained and has been modeled to have a radius of $\sim$30 AU, and potentially being surrounded by a halo \citep{verhoeff2011, casassus2013}. The inner disk is highly crystalline and is responsible for the mid-infrared emission \citep{leinert2004, vanboekel2005}. The outer disk is responsible for most of the far-IR emission and optically thick at near- and mid-infrared wavelengths. It is highly structured and asymmetric and has been studied in scattered light in the near-IR $H$-band, $K_{\rm s}$-band, and $L$-band \citep{fukagawa2006, rameau2012, casassus2012}. Recently, a PDI study by \citet{canovas2013} revealed the outer disk in polarized light and also showed the outer regions of the gap, without being able to detect the inner disk.

HD142527 has an extraordinarily high infrared flux compared to the stellar flux \citep[$F_{\rm IR}$/$F_{\star}$ = 0.92,][]{dominik2003}, which has been explained by \citet{verhoeff2011} with an unusually large scale height of the outer disk. Planetary-mass companions have been suggested multiple times to explain the eccentric shape of the outer disk, but direct searches for such companions using angular differential imaging have yielded null results \citep{rameau2012, casassus2013}. However, \citet{biller2012} suggest a stellar companion (0.1-0.4 M$_\sun$ at $\sim$0.088$\arcsec$) based on Sparse Aperture Masking observations. The validity of this claim is unclear \citep{casassus2013}. A recent study of the system with ALMA \citep{casassus2013} suggests that material from the outer disk is falling onto the inner disk via streamers through the gap, consistent with measured accretion rates onto the star \citep{lopez2006}. It also reveals a large asymmetry in the sub-mm continuum flux, possibly produced by a Rossby wave instability like the one shown for IRS 48 by \citet{marel2013}. The ALMA radial velocity data for HD142527, combined with the assumption that the spiral arms are trailing, also supports the suggestion by \citet{fujiwara2006} that the eastern side is the far side of the disk. 

The parameters of the HD142527 system are summarized in Table \ref{table:HD142527}. We adopt a value of 145 pc for the distance to the star because it is likely a member of the Upper Sco association \citep{acke2004b, zeeuw1999}. The value of 145 pc is well within the 3$\sigma$ error margin of the \emph{Hipparcos} measurement \citep{leeuwen2007}.

\begin{deluxetable*}{cc@{\hspace{10pt}$\times$}c@{$\times$}c@{$\times$}c@{=}cccccccc}
\tabletypesize{\footnotesize}
\centering
\tablecaption{Summary of observations. 
\label{table:observations}}           %
\tablewidth{450px}
\tablehead{
\colhead{}    &  \multicolumn{5}{c}{Integration Time} & \colhead{} &  \multicolumn{3}{c}{Observing Conditions} \\ 
\cline{2-6} \cline{8-11} \\ 
\colhead{Filter} & \colhead{DIT\tablenotemark{a}} & \colhead{NDIT\tablenotemark{a}} & \colhead{NINT\tablenotemark{a}} & \colhead{NPOS\tablenotemark{a}} & \colhead{Total\tablenotemark{a}} & \colhead{} & \colhead{Airmass} & \colhead{Seeing\tablenotemark{b}} & \colhead{$\tau_0$\tablenotemark{c}} & \colhead{Enc. Energy\tablenotemark{d}}
}
\startdata
$NB 1.64$ & 0.5s&36&1 & 3 & 54s&&1.07&1.05$\arcsec$&2.4ms&43.6$\%$\\
$H$ & 0.3454s&115&3 & 3 & 357.5s&&1.06&0.87$\arcsec$&2.9ms&48.9$\%$\\
$NB 1.64$ & 0.3454s&15&1 & 3 & 15.5s&&1.05&0.94$\arcsec$&2.7ms&50.3$\%$\\
$H$ & 0.3454s&115&3 & 3 & 357.5s&&1.05&1.21$\arcsec$&2.1ms&42.9$\%$\\
$NB 1.64$ & 0.3454s&15&1 & 3 & 15.5s&&1.06&1.53$\arcsec$&1.6ms&32.5$\%$\vspace{5pt}\\
$NB 2.17$ & 0.3454s&15&1 & 3 & 15.5s&&1.07&1.37$\arcsec$&1.7ms&35.8$\%$\\
$K_{\rm s}$ & 0.3454s&120&3 & 3 & 373s&&1.08&1.27$\arcsec$&2.2ms&40.7$\%$\\
$NB 2.17$ & 0.3454s&15&1 & 3 & 15.5s&&1.11&1.16$\arcsec$&2.2ms&41.5$\%$\\
$K_{\rm s}$ & 0.3454s&120&3 & 3 & 373s&&1.13&1.15$\arcsec$&2.2ms&39.7$\%$\\
$NB 2.17$ & 0.3454s&15&1 & 3 & 15.5s&&1.17&1.01$\arcsec$&2.5ms&37.6$\%$\\
\enddata
\tablenotetext{a}{The detector integration time (DIT) multiplied by the number of integrations per frame (NDIT) multiplied by the number of integrations per dither position (NINT) multiplied by the number of dither positions (NPOS) gives the total integration time per retarder plate position.}
\tablenotetext{b}{Average DIMM seeing in the optical during the observations, monitored by the seeing monitor at VLT.}
\tablenotetext{c}{Average coherence time of the atmosphere as calculated by the real time computer of the AO system.}
\tablenotetext{d}{Average encircled energy according to the ESO real time computer.}
\end{deluxetable*}

\section{Observations and data reduction}\label{observations_section}
The observations were performed on the night of July 23rd, 2012 with the NAOS/CONICA (NACO) instrument of the Very Large Telescope (VLT) at Cerro Paranal, Chile, in the $H$ and $K_{\rm s}$ filter. Because HD142527 is bright enough to saturate the detector at the shortest possible integration time (0.3454s), additional unsaturated images were taken in the narrow-band $NB1.64$ and $NB2.17$ filters.The total on-source integration times were 2860s and 2984s, respectively, in the $H$ and $K_{\rm s}$ filters, and 340s / 187s in the respective narrow-band filters. We did not observe any calibration stars, but use the central star itself for calibration. The observing conditions were mostly favorable with an average seeing in the optical of 1.04$\arcsec$ during the $H$ band observations and 1.21$\arcsec$ during the $K_{\rm s}$ band observations. The coherence time was 2ms or longer. The conditions during the narrow-band filter observations were slightly worse. Sky conditions were photometric and the airmass did not exceed 1.17. A summary of the observations is given in Table \ref{table:observations}.

We used the SL27 camera (pixel scale of 27 mas pixel$^{-1}$) in \emph{HighDynamic} mode and read out in \emph{Double RdRstRd} mode. The Wollaston prism splits the beam into an ordinary and extraordinary beam separated by 3.5$\arcsec$ on the detector. A polarimetric mask prevents the two beams from interfering, but limits the field of view to stripes of $\sim$27$\arcsec\times$3$\arcsec$. The rotatable half-wave retarder plate (HWP) controlling the orientation of the polarization was set to 0$^\circ$ / $-$45$^\circ$ to measure Stokes $Q$ and $-$22.5$^\circ$ / $-$67.5$^\circ$ to measure Stokes $U$. This means that we cycled through four retarder plate positions for each dither position and each integration.

The data reduction procedures, including corrections for instrumental effects, are described in detail in the appendix. %

We calculate the fractional polarizations $p_{\rm q}$ and $p_u$ using the double ratio method \citep[see Appendix or][]{quanz2011b} and obtain the Stokes $Q$ and $U$ parameters by multiplying with the intensity. Saturated parts of the images are mapped out. After that, the $Q$ and $U$ images from all dither positions are averaged.

Instead of calculating the polarization signal $P$ via sum of squares, i.e.
\begin{equation}
\label{equation_P}
P = \sqrt{Q^2 + U^2}
\end{equation}
we compute the tangential and radial Stokes parameters $P_\perp$ and $P_\parallel$ as
\begin{equation}
P_\perp=+Q\,{\rm cos}\,2\phi+U\,{\rm sin}\,2\phi
\label{equationFirst}
\end{equation}
\begin{equation}
P_\parallel=-Q\,{\rm sin}\,2\phi+U\,{\rm cos}\,2\phi
\end{equation}
where
\begin{equation}
\phi ={\rm arctan} \frac{x-x_0}{y-y_0}+\theta
\label{equationTheta}
\end{equation}
is the angle between the line from the star (at position ($x_0$, $y_0$)) to the location of interest and the sensor up direction. The (small) $\theta$ offset is needed due to a possibly not perfect alignment of the HWP \citep{witzel2010} and/or crosstalk effects and is determined from the data (see Appendix). Assuming that the polarized flux only has a tangential component, $P_\perp$ is equivalent to $P$ but unbiased (because of the sum-of-squares in Equation \ref{equation_P}, $P$ is always positive and noise will artificially increase the signal). $P_\parallel$ in this case contains no signal and can naturally provide an estimate of the noise levels. We estimate the error in $P_\perp$ as
\begin{equation}
\Delta P_\perp = \sqrt{\sigma^2_{P_\parallel}} / \sqrt{n_{\rm res}}\quad,
\end{equation}
where $\Delta P_\perp$ is the 1$\sigma$ uncertainty in $P_\perp$, $\sigma^2_{P_\parallel}$ is the variance in the $P_\parallel$ image in the region of interest and $n_{\rm res}$ is the number of resolution elements in the region of interest.

\begin{figure*}
\centering
\includegraphics[width=18cm]{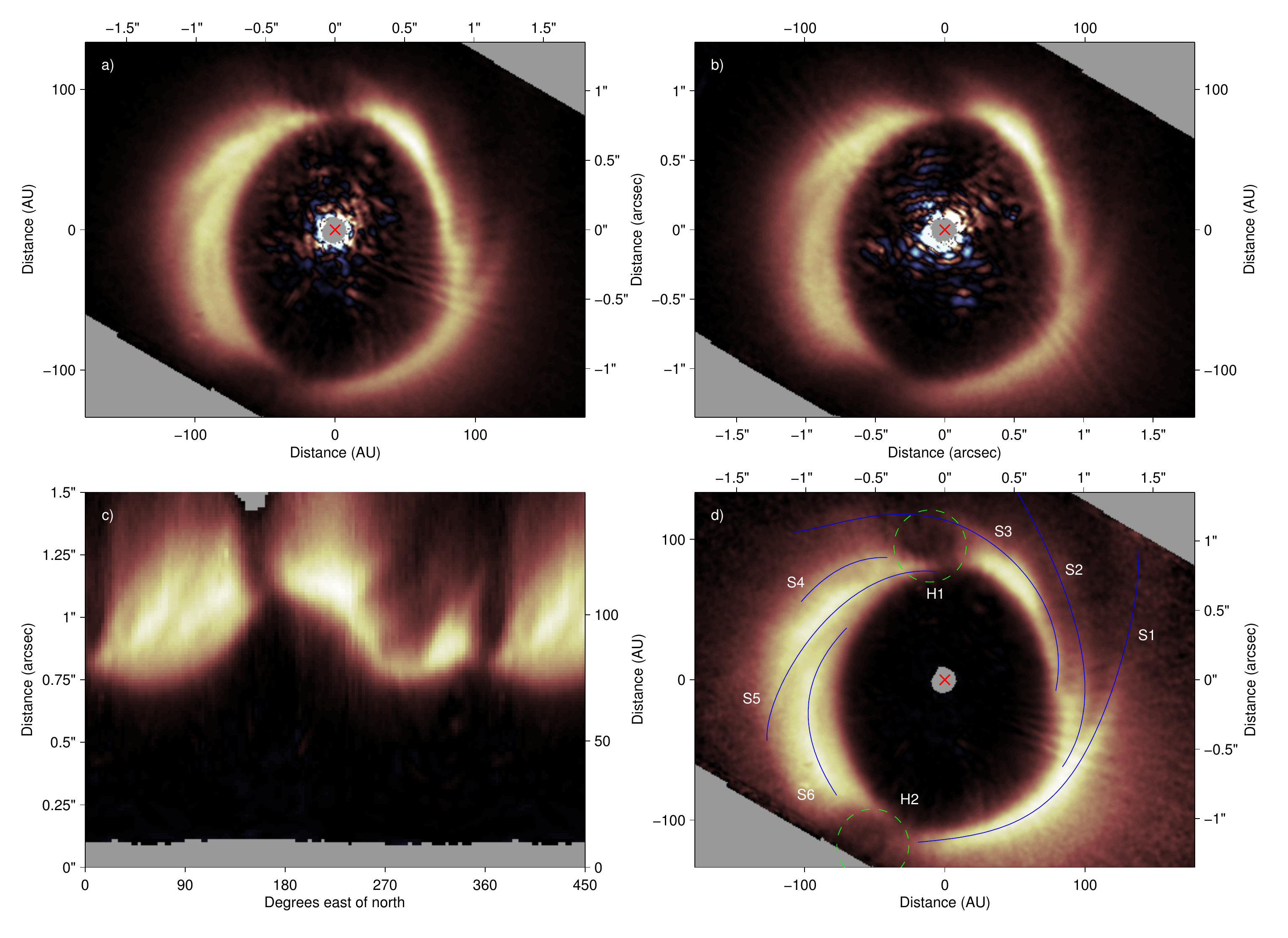}
\caption{NACO/PDI data of HD142527 in the $H$ and $K_{\rm s}$ band. All images in linear stretch. Because $P_\perp$ and $P_\parallel$ can be negative by construction, we show positive values in   and negative values in blue. Plots a) and b) show the final, reduced $H$ and $K_{\rm s}$ band images. North is up and east is to the left. Areas where no data is available due to saturation effects or the polarimetric mask are marked in grey. The red cross marks the position of the star. A radial mapping of the H band data is shown in c). Note that the data is plotted from 0 to 450$^\circ$ in order to show the hole in the disk at $\sim$0$^{\circ}$/360$^{\circ}$. In d), we mark the features seen in the disk. Spiral arms in the disk are marked S1 through S6. Two holes in the disk can be seen in the north (H1) and in the southeast (H2) at position angles of $\sim$0$^\circ$ and $\sim$160$^\circ$. The two small dots near S6 are effects from the $H$ band filter and not seen in the $K_{\rm s}$ image. There seems to be a kink in the disk in western direction at the starting point of the S3 spiral arm (seen in both filters). Images c) and d) have been scaled by $r^2$, i.e. the distance to the central star squared, to compensate for the drop-off in illumination from the star for this nearly face-on disk and to better bring out structures in the disk, while no scaling has been applied for a) and b).
\label{images}}
\end{figure*}

While we mostly use the broad-band images for our analysis as they go deeper and have higher signal-to-noise ratio (SNR), we can not use them to calibrate the surface brightness of the disk because the central star is saturated. For this, we use the unsaturated narrow-band images. Assuming that the magnitude in the broad- and narrow-band is approximately the same and further assuming that it is equal to the 2MASS magnitude in the respective band, we can get a zero-point for the magnitude in our broad-band images \citep[for a more detailed description, see][]{quanz2011b}. There are no signs of a Br$\gamma$ emission line or strong near-IR variability which would affect this calibration \citep{lopez2006}. We estimate that our absolute flux calibration is good to $\sim$30\%. We also use the narrow-band images for an estimate of the resulution achieved, since we cannot do this from the saturated broad-band images. The angular resolutions (FWHM) achieved in the $NB1.64$ and $NB2.17$ filters are 74 mas and 81 mas, respectively.

\section{Results}\label{results}

Figure~\ref{images} shows the $P_\perp$ images obtained by our reduction along with a polar coordinate mapping of our results. The disk is clearly seen at high signal-to-noise ratio (SNR) in both filters in $P_\perp$. The $P_\parallel$ images show mostly noise and faint residuals that are different between the two filters at the position of the disk. They do not show any clear or consistent structure and are not shown here (an example can be found in the appendix). The disk is clearly not symmetric, as already previously observed \citep[e.g.][]{fukagawa2006, rameau2012, canovas2013}, but shows a large, asymmetric (roughly elliptical) ring at $\sim100-150$ AU and a huge, nearly empty (see discussion in Section \ref{section:innerhole}) gap down to the inner working angle of 0.1$\arcsec$ / $\sim$15 AU.%

\begin{figure}
\centering
\includegraphics[height=13cm]{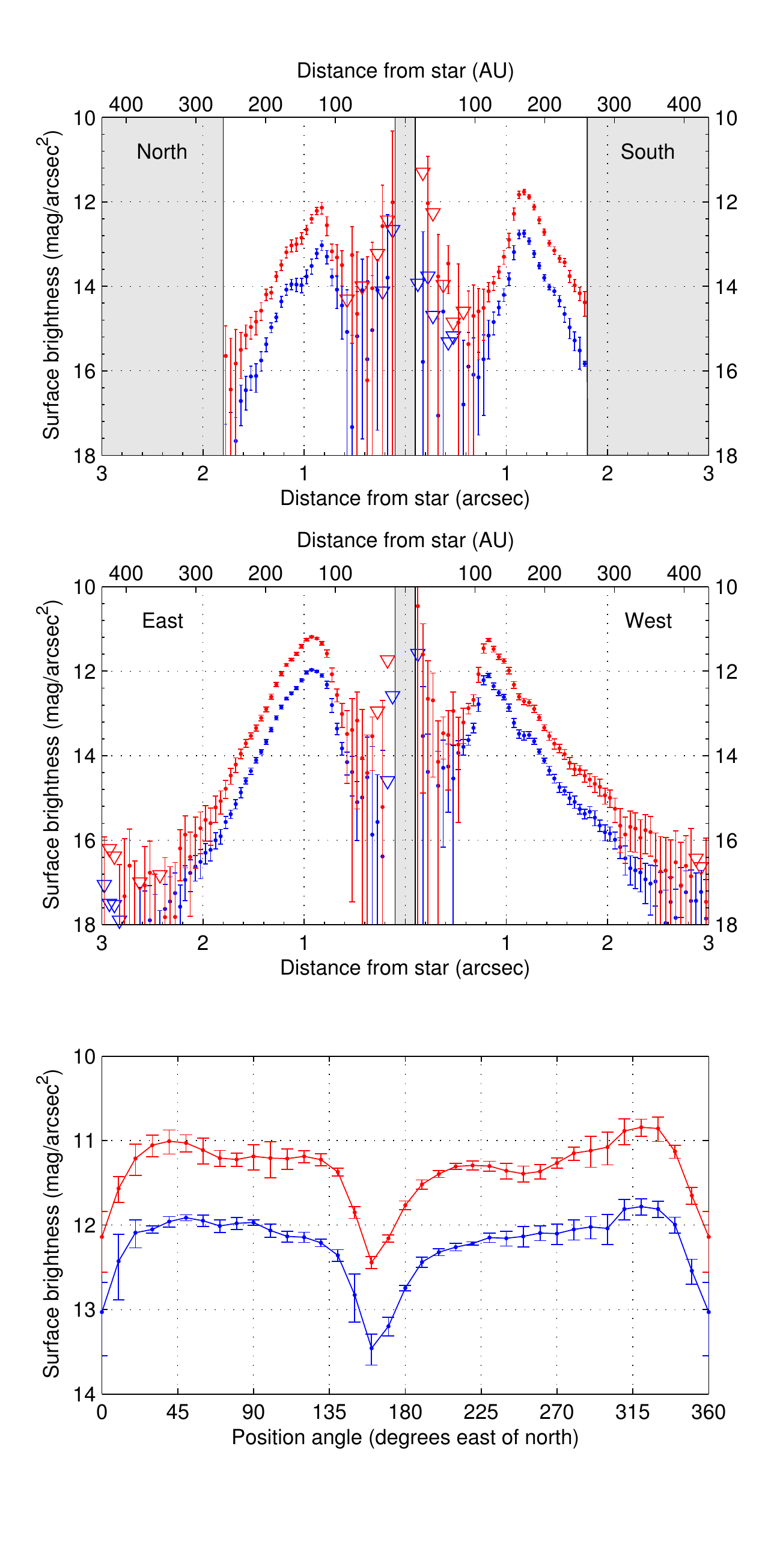}
\caption{Surface brightness of HD142527 in polarized scattered light in the $H$ band (blue) and the $K_{\rm s}$ band (red). Surface brightnesses and 1-$\sigma$ errors have been calculated as described in \citet{quanz2013} and in the appendix. Top two panels: Surface brightness in the North, East, South and West directions, calculated in radial steps of 0.05$\arcsec$ and azimuthal wedges of 10$^\circ$. Areas where no data is available (due to saturation or the polarimetric mask) are marked in grey. Downward-facing triangles represent 1-$\sigma$ upper limits. In the inner part, the signal-to-noise level is very low. The weak positive signal is however not just a bias, because as described in the appendix, our method of calculating the polarimetric flux has no known biases. For a deeper discussion of the flux in the gap, see section \ref{section:innerhole}. Bottom panel: The peak brightness of the ring (brightest point in given direction between 0.5$\arcsec$ and 1.5$\arcsec$) at the different position angles. Clearly visible are the two holes at PA $\sim$0$^\circ$ and $\sim$160$^\circ$. A systematic calibration error of up to $\sim$30$\%$ \citep[see][]{quanz2013} is not included in the error bars.}
\label{fig:SB}
\end{figure}

The disk itself displays a complex spiral arm structure, which we trace in Figure \ref{images} d). Two of these spiral arms (S1 and S2) have been detected before \citep{fukagawa2006, casassus2012}, S5 has also recently been seen by \citet{canovas2013}. The others are new detections. We are able to resolve the eastern side of the disk much more clearly, which also shows spiral structures. %
In the north (PA $\sim$0$^\circ$) and southeast (PA $\sim$160$^\circ$), significant holes (lack of polarized flux) are seen in the disk. As can be seen in Figure \ref{images} c), these holes seem to extend outwards, almost as if they were casting shadows. Interestingly, there is one spiral arm (S3) which passes through the northern hole and re-appears on the other side of the hole. These holes have been hinted at before \citep{casassus2012, canovas2013}, but not been seen at this resolution. %

In figure \ref{fig:SB}, we trace the surface brightness of the disk in the North, East, South and West direction. The disk shows a similar brightness in the eastern and western direction in polarized scattered light, in contrast to unpolarized scattered light, where the western side is brighter \citep{fukagawa2006}. The disk reaches its peak brightness of $\sim$11.7$\frac{mag}{arcsec^2}$ in $H$ band and $\sim$10.8$\frac{mag}{arcsec^2}$ in $K_{\rm s}$ band in the northwest, at a position angle of $\sim$330$^\circ$. Compared to the 2MASS $H-K_{\rm s}$ color of the star, the scattering is slightly red (by $\sim$0.16 magnitudes). Further out, the surface brightness drops rapidly. Fitting power-laws to the outer parts of the disc (outside 1.2$\arcsec$) leads to the following results for the power-law exponent in $H$ / $K_{\rm s}$ band: 
North: $-8.6\pm1.2$ / $-6.2\pm0.7$; South: $-7.0\pm0.6$ / $-6.1\pm0.3$; East: $-6.8\pm0.2$ / $-6.6\pm0.4$; West: $-4.5\pm0.2$ / $-4.5\pm0.2$. It is worth noting that starting from 1.2$\arcsec$ usually ignores the inner spiral arms. It also ignores the northern hole, which strongly affects the surface brightness in the northern direction between 0.8$\arcsec$ and 1.2$\arcsec$. The surface brightness profiles are somewhat steeper in the $H$ band compared to the $K_{\rm s}$ band, especially in the northern and southern direction. However, this result has to be taken with care as the number of points usable for the fit in the northern and southern direction is limited by the polarimetric mask.

\begin{figure*}
\centering
\includegraphics[width=18cm]{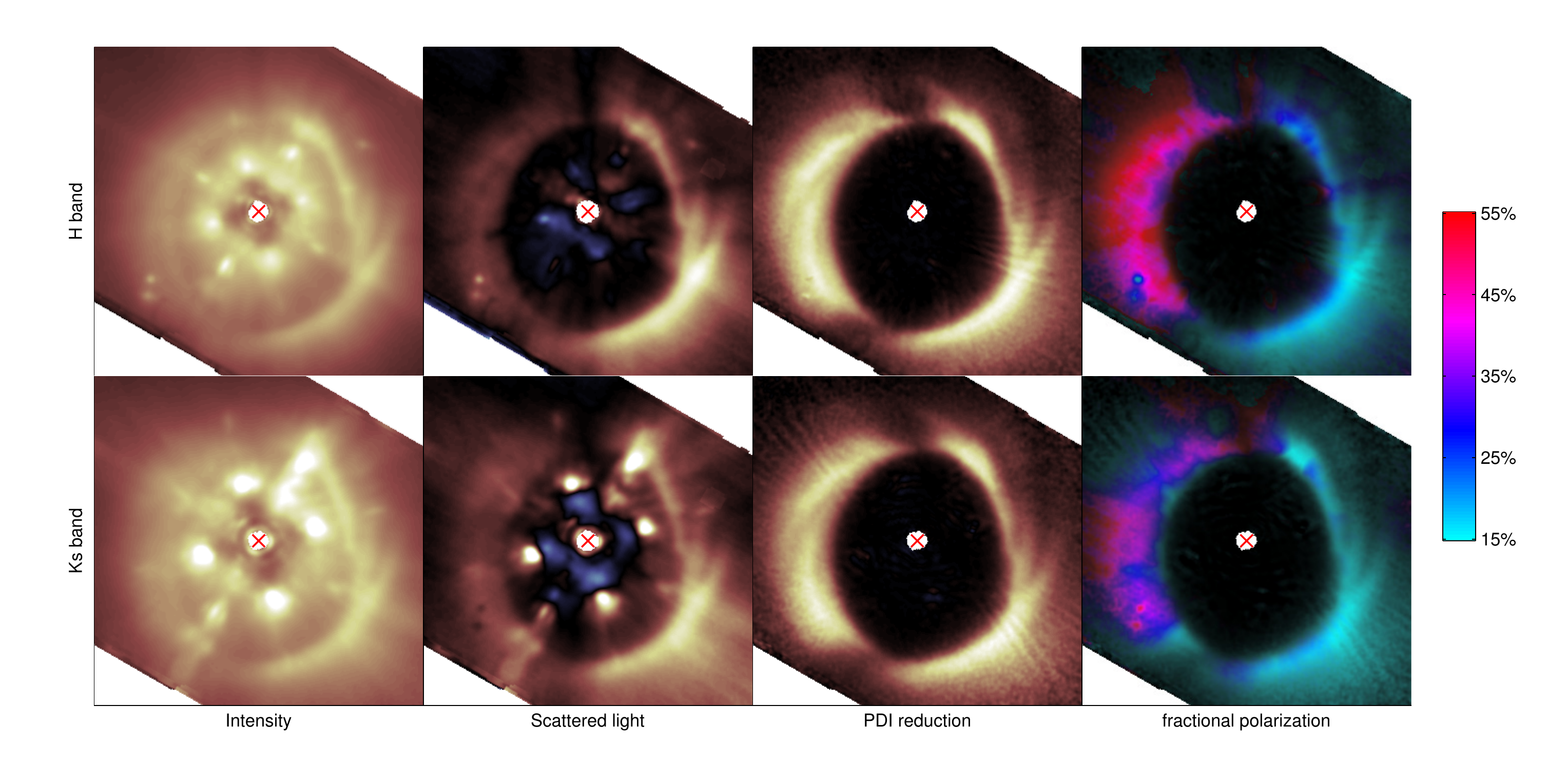}
\caption{Determination of fractional polarization in $H$ and $K_{\rm s}$ band. $H$ band in upper, $K_{\rm s}$ band in lower row. From left to right: Intensity image, intensity image after PSF subtraction, polarization image for comparison and fractional polarization. The PSF subtraction works significantly better in $H$ band because our reference stars (see text) were only taken in $H$ band. Furthermore, the bright spots produced by the adaptive optics lie further from the star and thus closer to the disk in $K_{\rm s}$ band. The fractional polarization is significantly higher in the eastern (backscattering) part of the disk than in the western (forward-scattering) part. Notice that the spiral structures seen in polarization are also present in the $H$ band scattered light image. All images scaled by $r^2$.
\label{fig:polFrac}}
\end{figure*}

\begin{figure}
\centering
\plotone{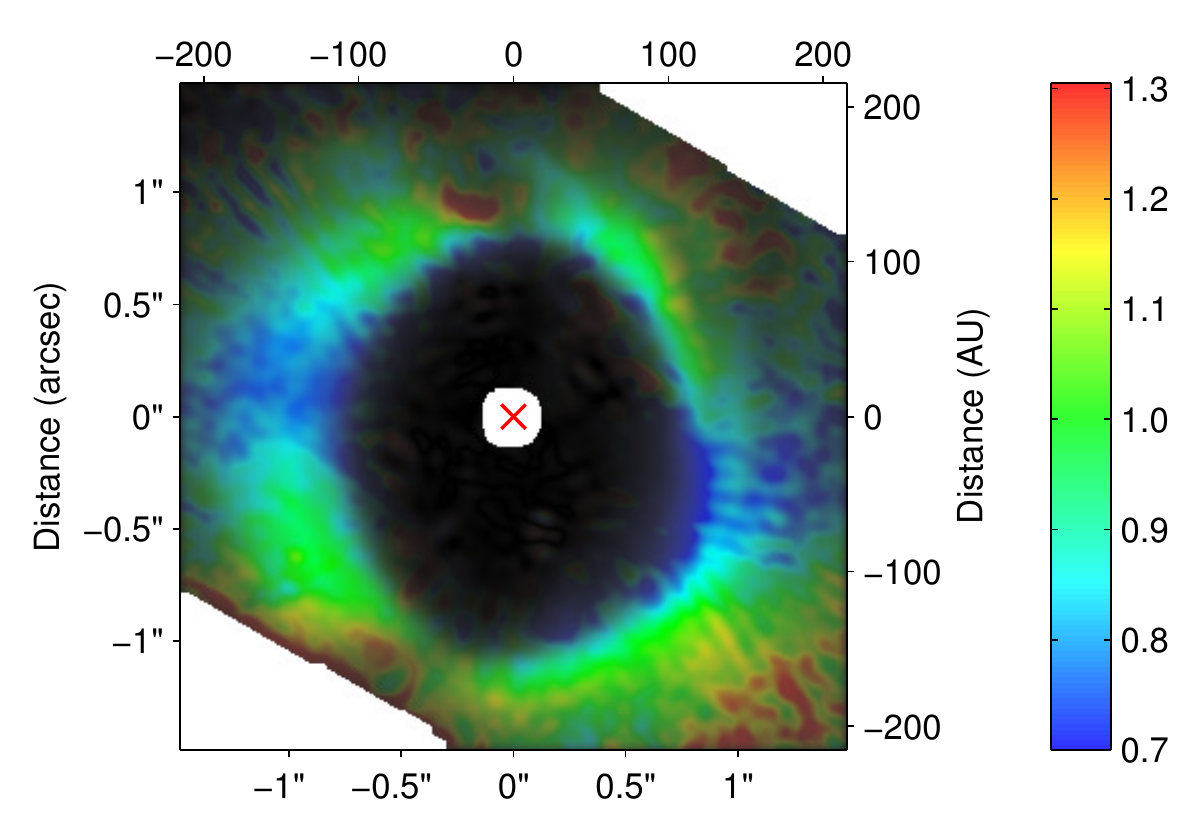}
\caption{Color of the disk in polarized light, displayed as [$K_{\rm s}$] - [$H$]. The disk is more red in the northern and southern direction than along the East-West axis. The hole in the northern direction appears to be particularly red compared to its surroundings. The data has been smoothed with a gaussian kernel of 1 detector pixel ($\sim$0.027$\arcsec$) width.
\label{fig:diskColor}}
\end{figure}

To measure the fractional polarization of the disk, we need a direct measure of the total scattered light. In the case of HD142527, the scattering from the disk is so strong that it can be seen directly in the intensity images we obtained, without a coronagraph or PSF subtraction - even in the individual frames. Because of this, we extract the information about the (non-polarized) scattered light directly from our data, rather than taking literature information, because we are then using the same dataset taken at the same time, which is less prone to errors.

During our observation run, we did not only observe HD169142 \citep{quanz2013}, SAO206462 \citep{garufi2013} and HD142527 (this paper), but also HD141569 and HD163296. Both these objects are known to harbor disks \citep[e.g.][]{weinberger1999, grady2000}, which however we did not or just barely detect, meaning that we essentially see the point spread function (PSF) of the star. The disk of HD169142 is rather weak, and the intensity image is also dominated by the PSF of the star, the directly measured polarization does not exceed 2$\%$ anywhere in the image. Furthermore, it is fairly symmetric in shape.

This means that we have three PSF reference stars at our disposal, and in our case we have the additional knowledge that the inner part of the disk (out to $\sim0.6$\arcsec) is nearly empty. Thus, by subtracting a linear combination of the reference star intensity images, we try to minimize the flux in this inner region. We use the HD169142, HD163296 and HD141569 $H$ band images for the $H$ band subtraction, and the HD169142 and HD141569 $H$ band images for the $K_{\rm s}$ band subtraction (we did not obtain $K_{\rm s}$ band images for these, and the HD163269 $K_{\rm s}$ band image is a very poor fit). We emphasize that this is not the ideal way to perform a PSF subtraction, especially because the other stars are known to harbor disks, but we opt for this way because we can use the same data for both the intensity and the PDI images.

We show our results in figure \ref{fig:polFrac}. As can be seen, the subtraction works significantly better in the $H$ band than in the $K_{\rm s}$ band. The measured polarization fractions are between $\sim$20$\%$ in the western (forward-scattering) part and $\sim$45$\%$ in the eastern (backward-scattering) part of the disk.

The intensity and polarization fraction images are plagued by various artifacts. For example, the spider pattern gets incorrectly over-subtracted, especially in the northern hole region, thus we deem the polarization fraction measurement not reliable in this region. There are several more artifacts seen, which do not represent real structures in the disk. We also emphasize that while there is a difference seen between the $H$ and $K_{\rm s}$ image, we do not think that this is significant, but is likely a result of the bad subtraction in the $K_{\rm s}$ band. The difference between the eastern and western sides of the disk, however, is significant and consistently detected in both filters.

We also compare the flux measured in the $H$ and $K_{\rm s}$ bands to determine the color of the disk. In general, the disk is slightly red: The total flux of reflected polarized light compared to the stellar flux, $\frac{F_{\rm pol}}{F_\star}$ within our field of view is (7.1$\pm$0.3)$\cdot10^{-3}$ in $H$ band and (8.2$\pm$0.3)$\cdot10^{-3}$ in $K_{\rm s}$ band, hinting towards relatively large grains \citep[c.f.][]{mulders2013}. 
The color of the disk is not constant, though: In the eastern and western part it appears to be more blue than in the north and south (figure \ref{fig:diskColor}). It is unlikely, though not impossible, that this pattern stems from the data reduction. A problem in the data reduction would likely manifest in an 8-leaf pattern (4 positive, 4 negative leafs) rather than in a 4-leaf (2 positive, 2 negative) pattern as is seen here. We also note that the disk seems to be particularly red inside the northern hole region.

\section{Analysis}\label{analysis}

\begin{figure*}
\centering
\plotone{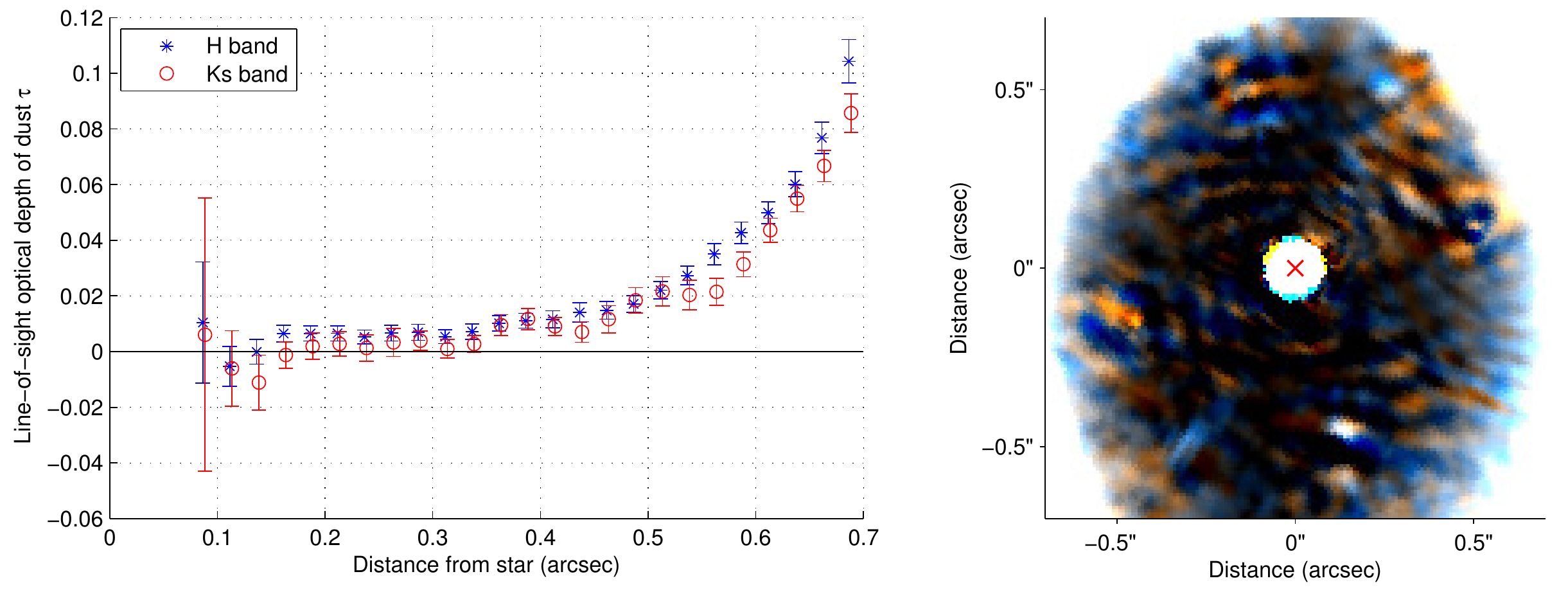}
\caption{Line-of-sight optical depth within the gap of HD142527, calculated as described in the text. Thus, these are upper limits. Left: Averaged optical depth $\tau$ (upper limits) inside 0.7\arcsec. The optical depth is very low, and there is no trace of the inner disk down to our inner working angle. Right: Color-composite image of the gap, scaled linearly from $\tau=0$ (black) to $\tau=0.1$ (white). H band represented in blue, Ks band represented in red.
\label{fig:innerHole}}
\end{figure*}

\subsection{Disk gap}
\label{section:innerhole}

It has been known for a while that between the massive, optically thick outer disk and the small inner disk, there is a large gap \citep{fukagawa2006, casassus2012}. In our images, the surface brightness drops dramatically inside the inner rim of the outer disk (c.f. figure \ref{fig:SB}). While just inside the outer disk, weak scattered light is still present, we do not detect scattered light at the 3$\sigma$ level inside of 0.4$\arcsec$ around the star. This also means that we do not detect any trace of the inner disk or halo of the system down to our inner working angle of 0.1\arcsec. 

To assess the amount of dust that can still be present in the gap below our detection limits, we would need to know the scattering properties of the dust grains, specifically their albedo and their polarization efficiency at the respective scattering angle. Because we do not want to rely on models, we use our data to derive lower limits for the product of the two (called $AP$ here), which we then turn into upper limits for the optical depth perpendicular to the disk plane. Because we take the outer, optically thick disk, where multiple scattering and self-shadowing effects will lower the albedo and the polarization efficiency and thus $AP$, this approach is conservative.

Assuming a dust particle of radius r at a distance of R from the star, the polarized flux due to scattering from the dust particle is:

\begin{equation}
F_{\rm d} = F_\star \frac{\pi r^2}{4\pi R^2} (AP)
\end{equation}

where $A$ is the albedo and $P$ is the polarization efficiency of the particle. Both depend on the scattering angle, but in our case the disk is close to face-on \citep[$\sim$20$^\circ$,][]{pontoppidan2011, verhoeff2011}, and we neglect inclination effects for this first-order estimate. In this form, this only holds true for single particles. However, if we add in the geometric fill factor of the dust (defined as $f_{\rm geo}=1-1/e^\tau$), we can use this formula for patches on the sky:

\begin{equation}
F_{\rm d} \leq F_\star \frac{S f_{\rm geo}}{4\pi R^2} (AP)
\end{equation}

were $S$ is the area of the patch on the sky we observe. We use the inequality because, as stated above, multiple scattering and self-shadowing effects can reduce the polarized flux. We can turn this into a calculation for $AP$:

and thus:

\begin{equation}
(AP) \geq \frac{4\pi R^2 F_{\rm d}}{F_\star S f_{\rm geo}}\
\end{equation}

We realize that the geometric filling factor cannot be larger than unity. Furthermore, the distance in the image plane provides a lower limit to the actual physical distance, and we use this formula to provide a lower limit for $AP$. To do so, we evaluate the formula above across the whole image frame and search for its maximum. The derived lower limits for $AP$ are 4.1\% in the $H$ band and 5.1\% in the $K_{\rm s}$ band, consistent with previously estimated albedos \citep[e.g.][]{mulders2013} and the polarization efficiencies  determined in section \ref{results}.

We turn these lower limits for the albedo and polarization efficiency into upper limits for the optical depth $\tau$ of the dust, by using the formulas above in reverse order. We then average over small annuli of 0.025\arcsec width. The results are shown, along with a scaled image of the gap, in Fig. \ref{fig:innerHole}. As can be seen, the optical depth in the inner part of the gap is very low, and compatible (on a 3-$\sigma$ level) with being zero. Further out, scattering can be seen, which could be due to dust drifting into the gap from the outer disk. This enhanced polarimetric flux in the outer regions of the gap is too strong to be explained by PSF smearing of the outer disk flux. Averaging over a larger annulus from 0.1\arcsec to 0.4\arcsec, we estimate the 3-$\sigma$ upper limit of the optical depth (averaged over the annulus) to be 0.012 in the $H$ and 0.011 in the $K_{\rm s}$ band. We do not see significant asymmetric structures in the gap. Especially, we do not detect any small-grain counterpart to the gas streamers seen by \citet{casassus2013}.

We need to mention at this point that a geometrically perfectly thin disk would not be detected in our observation, due to the incidence angle of the light. Furthermore, if the streamers have a small enough scale height, they could be shadowed by the (unseen) inner disk. However, we do not deem this scenario very realistic. A perfectly thin disk would be unphysical, and there is no reason to believe that the inner disk should be settled to very small scale heights while the outer disk has an extraordinarily large one.

\subsection{Disk model geometry}

The disk is known to be inclined, it is known to be eccentric and there is an estimate on the scale height of the inner rim \citep{pontoppidan2011, verhoeff2011}. Because we have such high-resolution imagery available, we can try to directly determine the relevant parameters of the disk. If we assume the inner hole to be elliptical in shape \citep[for example because of being carved out by an orbiting planet, c.f.][]{casassus2012} and the height of the inner rim to be constant, the inner rim of the disk can be described by six parameters: Two for the inclination (angle of inclination and position angle for the inclination), two for the eccentricity (eccentricity and position angle of eccentricity), one for the scale height of the inner rim and one for the semi-major axis. If inclination and scale height are small enough, as is the case for HD142527, we can see through the inner hole. The visual inner edge of the inner hole will not directly be the inner rim of the disk, but actually the edge of the inner rim - on one side, it is the front edge, on the other side, it is the rear edge.

\begin{deluxetable}{lcc}
\centering
\tablecaption{Parameters of geometric disk model
\label{table:diskParameters}}           %
\tablewidth{200px}
\tablehead{
\colhead{Parameter}&{from H band}&{from Ks band} \\
}
\startdata
inclination&$23^{+6}_{-8}$&$23^{+6}_{-9}$\vspace{2pt}\\
incl. PA&$-3^{+1}_{-2}$&$-2^{+2}_{-1}$\vspace{2pt}\\
eccentricity&$0.137^{+0.005}_{-0.005}$&$0.137^{+0.006}_{-0.005}$\vspace{2pt}\\
ecc. PA&$182^{+1}_{-2}$&$183^{+2}_{-1}$\vspace{2pt}\\
semimajor axis&$143^{+2}_{-3}$&$143^{+2}_{-2}$\vspace{2pt}\\
scale height&$42^{+51}_{-24}$&$37^{+58}_{-24}$\\
\enddata
\tablecomments{Parameters of the disk as derived from H and Ks band images via MCMC modeling. Position angles are given in degrees east of north, inclination is given in degrees, semimajor axis and scale height are given in AU.}
\end{deluxetable}

\begin{figure}
\centering
\plotone{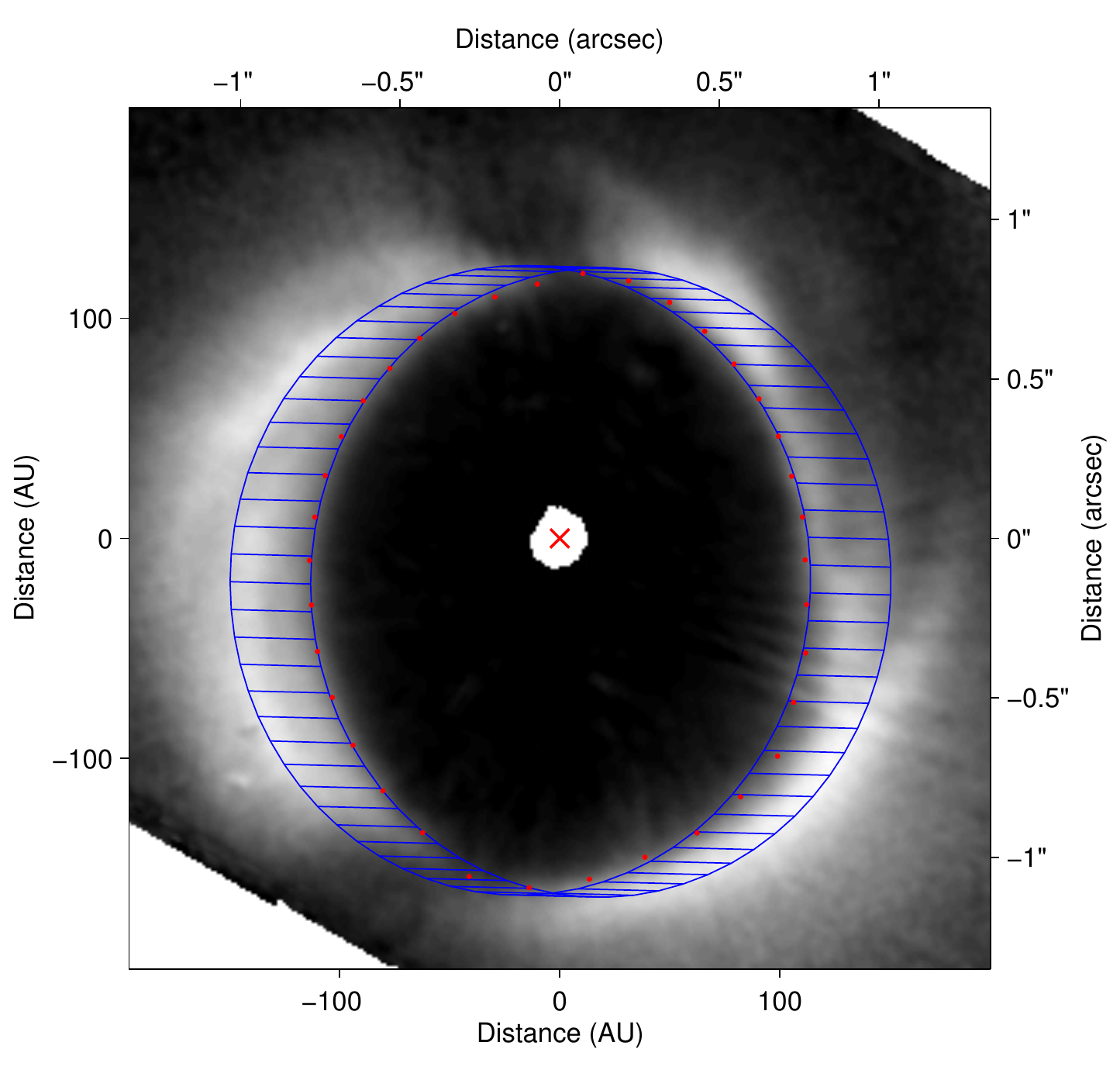}
\caption{The model inner rim acquired by MCMC fitting overlaid over the image of the disk ($H$ band data). The red dots mark our tracing of the inner rim, taking into account the smearing of the inner rim by the PSF.
\label{fig:modelDisk}}
\end{figure}

We trace the visual inner rim at 36 points (spaced 10$^\circ$ in position angle apart) using radial surface brightness profiles and a threshold of 30$\%$ w.r.t. the peak brightness of the ring in this direction. We also account for the PSF smearing by adding 0.04$\arcsec$ to the value determined in this way and assume the error on this measurement to be proportional to inverse of the local gradient of the radial surface brightness. Our trace of the inner rim can be seen in figure \ref{fig:modelDisk} (red dots).

We then generate a model inner rim which we cast onto the image. By doing this, we can compare the position of the visual inner rim with the one which the model disk would generate. We use the adaptive Markov-Chain Monte-Carlo (MCMC) implementation by M. Laine \citep[http://helios.fmi.fi/\textasciitilde lainema/mcmc/]{MCMCbook}  to marginalize over all six parameters, performing 10$^6$ realizations to properly sample the parameter space. By doing so, we can self-consistently obtain estimates for all six parameters.

\begin{figure}
\centering
\plotone{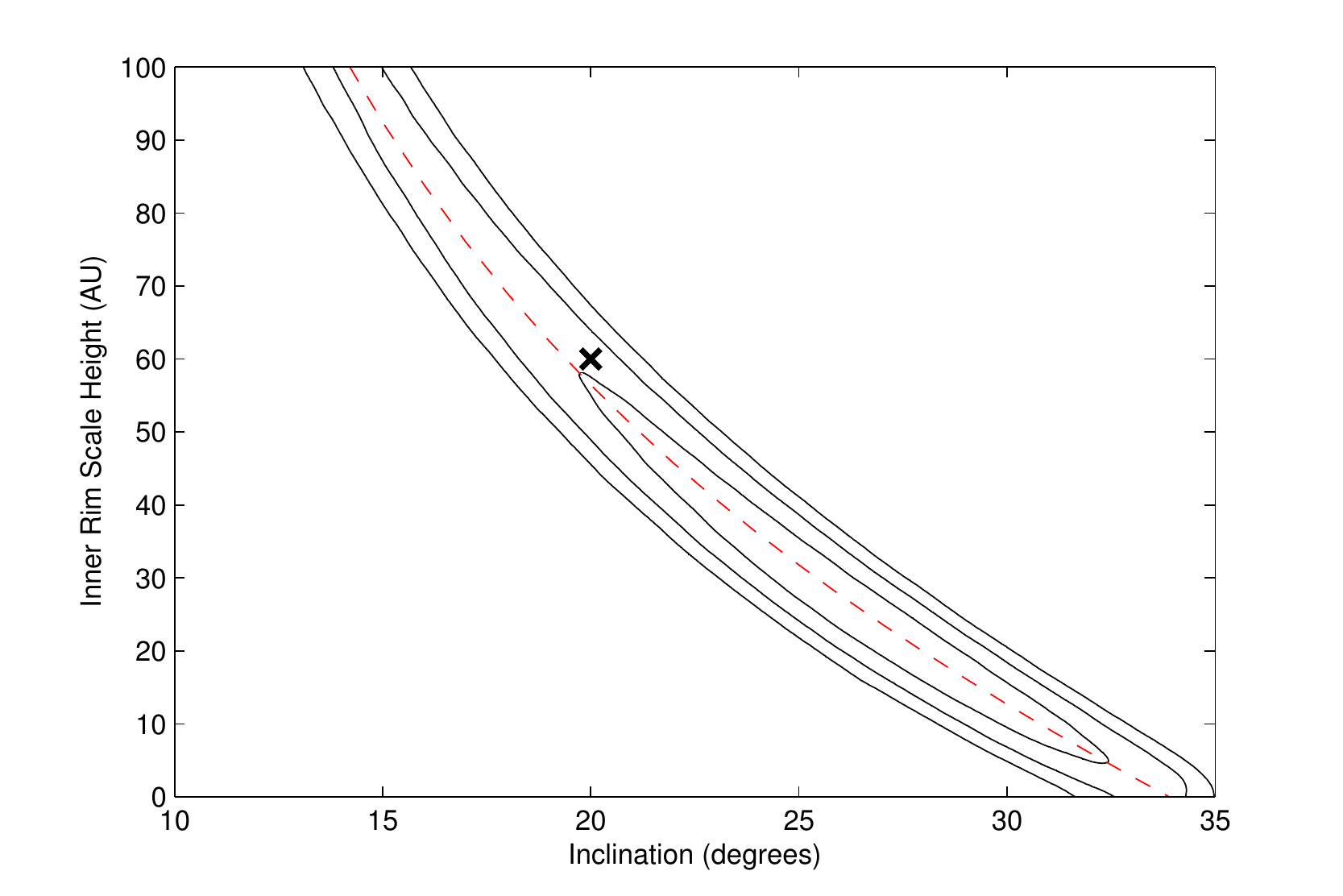}
\caption{Contour plot of the inclination and scale height parameters for our fit. The three contours (solid black lines) represent the 1-, 2-, and 3-$\sigma$ levels generated from combining the probability maps in the $H$ and $K_{\rm s}$ band. The dashed red line represents a functional dependence (not fitted) of $h=\frac{\rm cos(i)-0.83}{\rm sin(i)}*176 \rm AU$, where $i$ is the inclination and $h$ is the scale height. The black cross marks the position of $(i,h) = (20^\circ, 60 \rm AU)$, which are the values suggested by \citet{verhoeff2011}.}
\label{fig:MCMC_IvsH}
\end{figure}

We summarize our results in table \ref{table:diskParameters}. Most of the parameters are only weakly or not at all correlated. An important exception are the inclination and the scale height of the inner rim, which are also relatively poorly constrained in our model. However, there is a very strong correlation between them, as we show in figure \ref{fig:MCMC_IvsH}. This strong correlation would allow for a better observational constraint on the inner rim scale height given a good measurement of the inclination. For example, an inclination of 20$^\circ$ yields a scale height for the inner rim of $59\pm6$ AU ($H$ band) and $54\pm6$ ($K_{\rm s}$ band), respectively.

In figure \ref{fig:modelDisk}, we show the model inner rim overplotted over the $H$ band polarized scattered light data using the median values for the parameters. The general shape of the optical inner rim is traced very well, especially on the eastern side of the disk, which is known to be the far side from modeling, but also from rotation curves and assuming that the spiral arms are trailing spirals \citep{verhoeff2011, casassus2013}. As can be seen, on the eastern side, the inner rim is resolved, and the spiral structures seem to actually be present in the inner wall. Further structure is seen outside the inner wall. On the western side of the disk, the fit is less accurate, with the disk seemingly bulging outwards in the southwestern direction. One reason could be that we are not tracing the optical inner rim properly - there is a faint scattered light signature further in in this region. In any case, there seems to be some kind of asymmetry in the inner rim in the southwestern direction.

We note at this point that it is unlikely that the inner rim forms a perfect ellipse. It is also unlikely that the scale height is absolutely constant over the entire rim. Our simplified model can explain the visual signature of the inner rim remarkably well. However, the errors because of deviation of the real disk from our simplified model likely dominate over the statistical errors given above, especially for the parameters of eccentricity and semi-major axis.

\section{Discussion}\label{discussion}

With the data available from past observations and the new polarimetric differential imaging data from this paper, we can try to draw a consistent picture of the HD142527 system.

The system consists of an inner disk and/or halo close to the star and a massive outer disk with large scale height. While the large scale height of the outer disk has been explained in terms of hydrostatic equlibrium by \citet{verhoeff2011}, these authors also suggest that the halo and inner disk extend out to $\sim$30 AU. In our observations, we do not detect the inner disk at radii down to 0.1\arcsec. This means that the inner disk / halo probably does not extend this far out, but resides inside of $\sim$15 AU, because otherwise we should have detected it.%

We see that the gap is very devoid of dust. It is very likely that some dust is present in the gap, because we know that there is an optically thick CO disk filling the entire gap \citep{casassus2013}. This gas is likely to drag some dust along and into the gap, but this amount must be very small, as we see from our observations. We do not see any structure inside the gap. This means that we also do not see any streamers of small grains that might corresponding to gas falling onto the inner disk from the outer one, as discussed by \citet{casassus2013}. Our results in this respect are consistent with the results by \citet{canovas2013}, the only other attempt so far to detect scattered light at very small separations for HD142527.

There has to be some clearing mechanism that empties the gap. While photoevaporation can explain holes up to $\sim$20 AU in accreting and even larger holes for non-accreting disks \citep{owen2011}, it can not explain the gap in HD142527, which has an accretion rate of 6.9 $\cdot 10^{-8}$ ${\rm M}_\sun{\rm yr}^{-1}$ \citep{lopez2006} and a very large hole. Photoevaporation would also not explain why there should still be an inner disk, which would be expected to clear on a a viscous timescale of $t_{\rm visc} \sim 10^5$yr.
Dust grain growth is an unlikely explanation for the gap because the it is also seen in the sub-mm \citep{casassus2013}, meaning that the surface density of $\sim$mm-sized grains is low. %
\citet{chiang2007} show that MRI could grow an existing inner  gap, but \citet{dominik2011} point out that ongoing accretion would still pull significant amounts of dust into the gap, which we do not detect. Also, in this case, there still is the inner disk. 

\citet{biller2012} interpret asymmetries in sparse aperture masking observations in terms of a stellar companion at a distance of 88$\pm$5 mas ($\sim$13 AU) and with a mass of of 0.1-0.4 $M_\sun$. An alternative explanation for their measurements could be an asymmetry in the inner disk responsible for most of the near-IR flux \citep{casassus2013}. While our data do not allow us to confirm or refute either claim since they only reach in to $\sim$0.1$\arcsec$, we can discuss how a companion would affect the gap. According to \citet{artymowicz1994}, a binary can clear a circumbinary disk out to a maximum of $\sim$2-3 semimajor radii, with binaries closer in mass and high eccentricities producing larger gaps. Assuming a semimajor axis of 13 AU for the companion, this is incompatible with the semimajor axis of the inner rim of $\sim$140 AU. The claimed companion could only explain the radius of the gap in the unlikely case of a much larger semimajor axis and a highly eccentric orbit (e $\gtrsim$ 0.7) combined with having observed it near the pericenter. It is worth noting that in the case of GG Tau A, a young binary system with a derived semimajor axis of 32.4 AU, the binary alone can also not explain the large inner radius of the outer disk of $\sim$180 AU, even though the mass ratio ($\mu$=0.47) is more favorable for a large hole than in the case of HD142527 and the eccentricity of e=0.34 is substantial \citep{beust2005}.

Because of this, we deem the most likely explanation for the large gap to be one or more planetary-mass companions which clear out most of the gas and dust in this region. This also is in concordance with the relatively sharp inner edge of the disk. \citet{casassus2012} have shown that a single planet could be able to produce a large, eccentric cavity for the disk of HD142527. Their simulation assumes a planet of 10 $M_{\rm Jup}$, which is not consistent with the detection limits of \citet{rameau2012}, but \citet{hosseinbor2007} show that also lower mass planets can clear eccentric gaps. Such a planet is still feasible even if the proposed stellar companion by \citet{biller2012} is real. Using the derivations of \citet{holman1999} and assuming an eccentricity of e=0 and a mass ratio of $\mu$=0.1, planetary orbits should be stable outside of $\sim$25 AU ($\sim$0.18$\arcsec$).

The outer disk is known to be very asymmetric in the sub-mm tracing mm-sized particles. A large horseshoe is seen in the northern direction of the disk, while in the south, the thermal emission is depleted by a factor of $\sim$40 w.r.t. the north \citep{casassus2013}. While eccentric disks can show a certain degree of asymmetry due to a "traffic jam" effect at the pericenter, this factor is much too large to be explained by this effect. The enhancement is also at the apocenter rather than the pericenter of the disk, and can be much better explained by a vortex, which leads to a dust concentration at the azimuthal pressure maximum and facilitates grain growth and thus planet formation \citep{regaly2012, ataiee2013}.

In this context, the two holes in the disk in the north and southeast are very intriguing. They have also been seen by \citet{casassus2012} in the $L'$, $K_{\rm s}$ and $H_2$ bands. \citet{rameau2012} also see the two holes in $L'$ band. %
In both cases, the hole was not as clearly resolved. It was not seen as a circular structure, as in our images, but rather as a depletion in the ring of scattered light that could be seen. However, this makes it clear that the structure we see is not due to a depolarization effect, but it is also present in unpolarized scattered light. We also detect it in our PSF subtraction analysis (see figure \ref{fig:polFrac}). The location of the northern hole coincides with the location of the probable vortex. This could lead to two possible explanations: One where the small dust is depleted because of growth into larger dust particles, and the scattering at near-infrared wavelengths would be strongly suppressed. The other possible explanation is that a planet has already formed, and is causing the disk surface to bend inwards, and thus to lie in the shadow. \citet{jangcondell2009} has shown that this effect could be seen at near-infrared wavelengths, albeit at smaller separations. There is also an asymmetry in the mid-infrared thermal emission. \citet{verhoeff2011} see emission peaks at position angles of +/- 60$^\circ$ and speculate that a planetary-mass object at a P.A. of $\sim$0$^\circ$ could be responsible for a buildup of material at the Lagrange points. \citet{rameau2012} have estimated mass limits for planets that might be hidden in these both gaps (5 $M_{jup}$ in the northern and 4 $M_{jup}$ in the southeastern gap), so any companion in these would have to be smaller than this. %

We adopt a model of the disk with an eccentric, inclined inner rim and constant inner rim scale height. The parameters we estimate for this model are well constrained due to the high signal-to-noise ratio and the high resolution of our images. The two parameters not well constrained (inclination and scale height) are strongly correlated. If we compare our results to the results of \citet{verhoeff2011}, we see that they differ. First, they use an axi-symmetric model, i.e. zero eccentricity. Their model uses an inner radius of 130 AU and an outer radius of 200 AU. We determine a semi-major axis of 143$\pm$3 AU, but the disk is closer to the star at some points due to the eccentricity. We also want to emphasize that we can trace the disk to distances of more than 300 AU in the eastern and western direction (c.f. figure \ref{fig:SB}). The inclination and scale height they use (20$^\circ$, 60 AU) are marked in figure \ref{fig:MCMC_IvsH} and are consistent with our results at the ~1$\sigma$ level. We stress at this point that an accurate measurement of the inclination, for example from rotation curves, would help a lot to break the degeneracy between scale height and inclination. Unfortunately, neither \citet{verhoeff2011} nor \citet{casassus2013} provide an error estimate for their inclination determination of 20$^\circ$. \citet{pontoppidan2011} provide an error estimate of 20$\pm$2$^\circ$, but this is based on the assumption that $M_{\star}=3.5M_{\sun}$. If the true mass of the star is lower (c.f. table \ref{table:HD142527}), the inclination would be higher, and the respective scale height would be lower.

The polarization fraction of the disk varies strongly between the eastern and the western side. While it is around 20\% on the western side, it seems to be as high as 45\% in the east. Due to our reduction technique, we expect the error on the second number quoted to be high (the disk is faint and close in, the region is dominated by the stellar PSF). However, we can see that while the two sides are of comparable brightness in polarized reflected light, it has been seen before that the western side is significantly brighter in scattered light \citep{fukagawa2006}, a feature which we also detect. We conclude that in fact, the polarization fraction does vary strongly within the disk, with the eastern side being more strongly polarized. One reason for this could be the different scattering angles between the western side (which is closer to us) and the eastern side (which is further away). %
A fractional polarization of this magnitude and a discrepancy between the forward- and backward-facing side are not unusual for protoplanetary disks and has been observed in AB Aurigae before by \citet{perrin2009}. The difference between forward- and backward-scattering polarization efficiency is consistent with the scattering theory presented in their paper, so we conclude that this is a likely explanation, but detailed modeling in future work will be required to disentangle geometry effects from dust scattering properties. %

\section{Conclusion}\label{conclusion}

In this paper, we image the disk of HD142527 at unprecedented signal-to-noise ratio and resolution. We achieve an inner working angle of 0.1$\arcsec$ and trace the disk out to more than 300 AU in the eastern and western direction, identifying a large amount of substructure in the disk.%
We also infer the amount of scattered light via a crude PSF subtraction using other target stars in order to determine the fractional polarization of the scattered light and determine the color of the disk. We use an MCMC code to determine the basic parameters of the inner rim of the disk. 

Our key findings are:

\begin{enumerate}

\item{There is a large amount of structure present in the outer disk. We count at least six spiral arms, and there are two prominent holes seen in the disk.}

\item{The gap between the (unseen) inner disk and the massive outer disk is large and mostly empty. We detect only faint, if any, scattering in the inner region and demonstrate the perpendicular optical depth to be at most $\sim$ 0.01. We also do not detect any small-grain correspondence to the gas "streamers" suggested by \citet{casassus2013}.}

\item{Using an adaptive MCMC code, we determine six basic parameters for the inner rim of the outer disk: Inclination (2 parameters), eccentricity (2 parameters), semi-major axis and scale height. Most of these are very well constrained, with the exception of scale height and inclination. These are very strongly correlated, and an improved measure of the inclination would improve our measurement of the scale height. Our results are compatible with existing models \citep{verhoeff2011}. For the first time, we measure the scale height of a disk directly from an image.}

\item{Our geometrical analysis shows that in the eastern (far) side of the disk, because of the very large scale height and the slight inclination, we are actually looking onto the inner wall of the disk.}

\item{The polarization efficiency in the disk is larger in the eastern (far) side of the disk than in the western (near) side. This is consistent with scattering theory and previous results \citep{canovas2013}, although we measure a significantly higher polarization fraction of $\sim 20-45\%$.}

\item{The average color of the disk is slightly red between the $H$ and $K_{\rm s}$ band. There seem to be small, so far unexplained color differences within the disk. Especially, the disk seems to be more red in the direction of the northern and southeastern hole.}

\end{enumerate}

We furthermore emphasize the hole in the north of the disk. This asymmetric structure lines up with asymmetric structures seen before in the mid-infrared and sub-mm wavelength ranges. A possible and intriguing explanation would be a forming or existing planet, another one would be strong grain growth due to a vortex instability. Future observations will be required to answer the question of the nature of this asymmetry.

\acknowledgments
This research has made use of the SIMBAD database, operated at CDS, Strasbourg, France. We thank the staff at VLT for their excellent support during the observations and F. Meru for useful discussions. This work is supported by the Swiss National Science Foundation.

{\it Facilities:} \facility{VLT:Yepun (NACO)}

\nocite{*}
\bibliographystyle{apj.bst}
\bibliography{HD142527_bibliography.bib}

\begin{appendix}

This appendix describes the data reduction techniques used to generate the scattered light images that appear in this paper as well as in the papers on HD 169142 \citep{quanz2013} and SAO 206462 \citep{garufi2013}, as well as possibly future papers.

\section{Image preparation}

All frames from the detector are dark frame subtracted and flatfielded in the standard way using darkframes of the appropriate exposure time and flatfields of the corresponding band, with all optical elements used for the observations in place. Unfortunately, the NACO detector shows a time-variable noise phenomenon across the detector rows, which can not be eliminated by the darkframing process. Because of this, we employ a row mean subtraction, where we calculate the mean in each row of each of the four 512x512 read-out electronics of the NACO detector (mapping out the star and the more noise-affected sensor boundaries) and subtract it from the entire row. Note that this also eliminates any sky background (which is in any case very low because of the short integration times and would also be eliminated in the PDI subtraction process, given it is unpolarized). After applying this process, the resulting images show only a very weak residual readout noise pattern.

From the darkframes and flatfields, bad pixel maps are constructed by mapping bad pixels (hot, dead or randomly varying). The value at the bad pixel is then replaced by the average of its surrounding pixels. The result is a clean, darkframed and flatfielded image which still contains the images of the ordinary and extraordinary beam.

We do not apply special care for ghosts in the images. These are usually unpolarized and cancel out in the PDI reduction process. We also tried to avoid having the ghosts lie on top of the ordinary or extraordinary beam in our observation plan by not putting the star close to the horizontal center of the detector, so that the (weak) ghosts lie outside the region of interest and do not cause any problems.

\section{Extraction and centering}

The ordinary and extraordinary beams are crudely extracted by finding the emission maxima and extracting the region around them. Then, the image peaks are determined by fitting a 2-dimensional gaussian function to the stellar PSF halo. Values above the sensor linearity limit (10000 counts) are not taken into account for this fitting procedure. We estimate that the accuracy with which we can estimate the stellar position is $\sim0.2$ pixels ($\sim0.005$\arcsec or $\sim0.8$ AU at a distance of 145 pc). The images are upscaled by a factor of 3 using bicubic interpolation and shifted to a common center using bilinear interpolation. The upscaling is done in order to minimize artificial and uncontrolled smoothing by the shifting process, which inevitably occurs when shifting images by fractional amounts of a pixel.

\section{Instrumental polarization correction}

The NACO instrument was not designed primarily with the focus on polarimetric observations. There are several inclined surfaces within the instrument which generate instrumental polarization, as described by \citet{witzel2010}. Because of this, we need to correct for instrumental polarization if we want to achieve the best possible signal-to-noise ratio and inner working angle for our observations. There are several instrumental effects that can occur and have to be taken into account. First, the ordinary and extraordinary beam have different transition efficiencies due to polarization effects in the light path in front of the Wollaston prism. Second, the rotation of the half-wave plate could be not perfectly aligned or instrumental polarization could slightly rotate the polarization direction. Third, the actual Stokes vectors (I, Q, U, V) suffer from polarization crosstalk, as shown by \citet{witzel2010}.

We try to address all these possible effects by estimating them from the data. The fundamental assumption we make is that the central star is unpolarized. However, even if this was not the case and the star was polarized either intrinsically or from interstellar polarization, our data reduction would still extract the scattered light from the disk, although to a lesser degree of accuracy. Only if the star was intrinsically polarized at a level which is significant compared to the polarization efficiency of the grains (usually $>10\%$), the extraction of the disk scattered light would break down, while interstellar polarization would affect both the star and the light scattered off its disk and would be removed by our calibration process. We do not expect our target stars to be highly polarized.

In order to correct for the first-order instrumental polarization, we impose that the flux in the ordinary and extraordinary beam should be the same. Unfortunately, we can not measure the flux of the star directly, because the central pixels in our $H$ and $K_{\rm s}$ band images are saturated at the core of the PSF. However, we know that also the halo of the star should be unpolarized. The stellar halo close to the star is much stronger than the scattered light from the disk. Furthermore, for an axisymmetric disk seen face-on, the integrated light in an annulus around the star should be unpolarized as well. Because of this, we chose to calculate the ratio $X_{\rm o/e} = \frac{f_{\rm o}}{f_{\rm e}}$ of the ordinary to extraordinary flux in an annulus between $R_{\rm in}$ and $R_{\rm out}$. We then multiply the image of the extraordinary beam by $(X_{\rm o/e})^{\frac{1}{2}}$ and the ordinary beam by $(X_{\rm o/e})^{-\frac{1}{2}}$ to equalize the flux, thus imposing a 0$\%$ polarization for the stellar halo.

\begin{figure}
\centering
\plotone{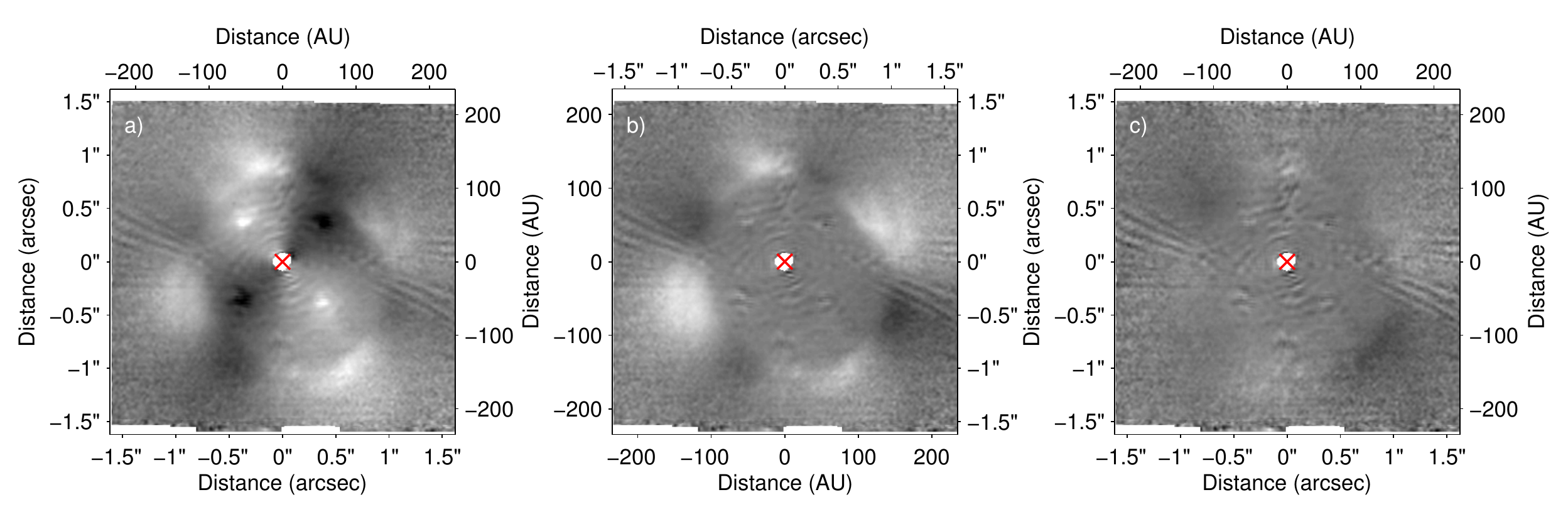}
\caption{Ur image before and after applying instrumental polarization correction. a) Ur image with no instrumental corrections at all. Strong residuals are seen, both in the disk and from the adaptive optics spots. b) After equalizing the ordinary and extraordinary beams. The adaptive optics spots are essentially gone, but there clearly is structure left where the disk is. c) After also applying the correction for U and the rotation correction. Most of the structure is gone, only faint residuals remain. All images are scaled in the same way, with zero mapped to median gray.}
\label{Fig_UrComparison}
\end{figure}

This improves the quality of the data significantly (see Fig. \ref{Fig_UrComparison} b)), but further instrumental polarization effects remain. Especially, there is crosstalk between the Stokes vectors. In all the disks we analyzed, the Stokes Q vector is stronger (more flux) than the Stokes U vector. In a symmetric disk, this is not expected, and from the analysis of all the disks we observe, we conclude that it is an instrumental effect which we then try to quantify and calibrate. Because it is not necessarily the case that the total flux in the Stokes Q and U vectors is the same for a non-symmetric disk, we need a method to calculate the suppression of flux in the Stokes U vector (i.e. crosstalk to circular polarization, see \citet{witzel2010}) other than a simple ratio of the Q/U fluxes. We estimate the efficiency of the measurement of Stokes $U$, denoted $e_{\rm U}$, in a self-consistent way from the data. Assuming that the polarization direction in the disk is tangential (centrally symmetric) and that noise can be neglected, we expect the number of pixels in an annulus around the star, if there is a disk, in which $|Q| > |U|$, to be equal to the number of pixels where $|U| > |Q|$. We thus take an annulus where the disk is clearly detected (high SNR) and calculate $e_{\rm U}$ such that $\sum\mathds{1}_{|Q|>|U/e_{\rm U}|}=\sum\mathds{1}_{|U/e_{\rm U}|>|Q|}$ for the sum over all the pixels in this annulus. This estimate is unbiased, although it is subject to noise, but the noise in this estimate is insignificant for our disks (the other noise sources are much stronger). We then multiply our Stokes $U$ measurement by $\frac{1}{e_{\rm U}}$ in order to compensate for the fact that the efficiency of the Stokes $U$ measurement is smaller than the efficiency for Stokes $Q$. Note that this effectively means that we assume the efficiency for Stokes $Q$ to be 1.

In the same step, we calculate the offset angle $\theta$ from equation \ref{equationTheta}. For this, we calculate the tangential polarization $P_\parallel$ as described in equations \ref{equationFirst} through \ref{equationTheta}, choosing $\theta$ such that $\sum{P_\parallel} = 0$, i.e. the average of $P_\parallel$ over all pixels in our chosen annulus is zero. This corrects for any artificial rotation of the polarization direction introduced by either misalignments of the retarder plate or crosstalk between the Stokes vectors.

\begin{deluxetable}{lcccccc}
\centering
\tablecaption{Reduction parameters. 
\label{table:ReductionParameters}}           %
\tablehead{
\colhead{Target} & \colhead{Band} & \colhead{$R_{\rm in}$} & \colhead{$R_{\rm out}$} & \colhead{$e_{\rm U}$}  & \colhead{theta}
& \colhead{Reference(s)}}
\startdata
\multirow{2}{*}{HD142527} 
& H&$0.1\arcsec$&$0.4\arcsec$&$0.631$&$-3.7^{\circ}$&\multirow{2}{*}{this paper}\\
&K$_{\rm s}$&$0.1\arcsec$&$0.4\arcsec$&$0.615$&$-3.7^{\circ}$\vspace{5pt}&\\

\multirow{2}{*}{SAO 206462} 
&H&$0.15\arcsec$&$0.35\arcsec$&$0.614$&$-3.7^{\circ}$&\multirow{2}{*}{Garufi et al. 2013}\\
& K$_{\rm s}$&$0.15\arcsec$&$0.35\arcsec$&$0.645$&$-3.7^{\circ}$\vspace{5pt}&\\

HD 169142
&H&$0.15\arcsec$&$0.25\arcsec$&$0.718$&$-7.0^{\circ}$\vspace{5pt}&{Quanz et al. 2013}\\

\multirow{7}{*}{HD100546}
&H (2006)&$0.3\arcsec$&$0.8\arcsec$&$0.563$&$-6.6^{\circ}$&\multirow{7}{*}{\begin{tabular}{@{}c@{}}Quanz et al. 2011 \\ Avenhaus et al. in prep.\end{tabular}}\\
&K$_{\rm s}$ (2006)&$0.3\arcsec$&$0.8\arcsec$&$0.561$&$-6.9^{\circ}$&\\
&H (2013 / 1)\tablenotemark{(a)}&$0.3\arcsec$&$0.8\arcsec$&$0.629$&$-7.0^{\circ}$&\\
&H (2013 / 2)\tablenotemark{(a)}&$0.3\arcsec$&$0.8\arcsec$&$0.778$&$-5.2^{\circ}$&\\
&K$_{\rm s}$ (2013)&$0.3\arcsec$&$0.8\arcsec$&$0.550$&$-4.6^{\circ}$&\\
&L (2013)&$0\arcsec$&$0.1\arcsec$&$0.736$&$0^{\circ}\tablenotemark{(b)}$&\\
&H (2013, cube mode)&$0.3\arcsec$&$0.8\arcsec$&$0.747$&$-6.0^{\circ}$&\\
&K$_{\rm s}$ (2013, cube mode)&$0.3\arcsec$&$0.8\arcsec$&$0.632$&$-3.9^{\circ}$&\\

\enddata
\tablenotetext{(a)}{ The derotator was rotated during the observations by 45$^\circ$ in order to test the behavior of the Stokes $Q$ and $U$ vectors, leading to two separate datasets for the $H$ band}
\tablenotetext{(b)}{ The noise in the $L$ band did not allow to calculate $\theta$, thus it was assumed to be zero.}
\end{deluxetable}

As described, we calculate the tangential and radial polarization vectors $P_\perp$ and $P_\parallel$ from the data. $P_\parallel$ is expected to be zero in the absence of noise, and to not show any clear structure. In Figure \ref{Fig_UrComparison}, we summarize the effects of our instrumental polarization correction for the example of HD142527 in the $K_{\rm s}$ band. As can be seen, the steps of our instrumental polarization correction significantly reduce the structure seen in $P_\parallel$. They also reduce the amount of noise in the inner part of the disk in the $P_\perp$ image (not shown here). Because of this, we conclude that the instrumental polarization correction helps suppress instrumental effects and increase the quality of the measurement of the polarized light ($P_\perp$ image).

The reduction parameters for HD142527, HD169142 and SAO206462 can be found in Table \ref{table:ReductionParameters}. We show the inner and outer radius adopted for the equalization of the ordinary and extraordinary beam as well as the derived efficiency of the Stokes $U$ measurement and the value of $\theta$.

\section{Sensor linearity mapping}

Because we saturate the core of the point spread function (PSF), we can not use the innermost region. The NACO detector responds linearly only up to $\sim$10.000 detector counts. We mark pixels that are in the non-linear regime of the detector as a first step, before we perform the instrumental polarization correction. Because the images are scaled and centered, the individual pixels in the final image can have contributions from both linear and non-linear pixels. In order to achieve the best possible inner working angle while still preventing our data from becoming unusable because of sensor non-linearity effects, we adopt the following scheme: We track for each individual pixel how much it is influenced by non-linear pixels. If the influence on the total value is higher than 10\%, we mark this pixel as unreliable and do not incorporate it in the calculation of the Stokes parameters. This means that close to the inner working angle, some pixels in the final image may be calculated from only a subset of the individual sets of four images (for the four HWP rotations) each. While this does decrease the signal-to-noise level, we consider it to be better than having no data at all.

\section{Calculation of Stokes parameters and stacking}

Because of being more robust and unbiased, we use the double ratio rather than the difference method to calculate the Stokes parameters $Q$ and $U$ \citep{tinbergen2005}. This means that we derive $Q$ and $U$ as:

$$p_{\rm q} = \frac{R_{\rm Q}-1}{R_{\rm Q}+1}\quad;\quad\quad p_{\rm u} = \frac{R_{\rm U}-1}{R_{\rm U}+1}$$

with

$$R_{\rm Q} = \sqrt{\frac{I_{\rm ord}^{0^\circ} / I_{\rm extra}^{0^\circ}}{I_{\rm ord}^{-45^\circ} / I_{\rm extra}^{-45^\circ}}}\quad;\quad\quad R_{\rm U} = \sqrt{\frac{I_{\rm ord}^{-22.5^\circ} / I_{\rm extra}^{-22.5^\circ}}{I_{\rm ord}^{-67.5^\circ} / I_{\rm extra}^{-67.5^\circ}}}$$

Here, the subscripts refer to either the ordinary or extraordinary beam and the superscripts refer to the angular position of the HWP. The Stokes $Q$ and $U$ parameters are then simply calculated as

$$Q = p_{\rm q} * I_{\rm Q}\quad;\quad\quad U = p_{\rm u} * I_{\rm U}$$

where

$$I_{\rm Q} = (I_{\rm ord}^{0^\circ} + I_{\rm extra}^{0^\circ} + I_{\rm ord}^{-45^\circ} + I_{\rm extra}^{-45^\circ})/2\quad;\quad\quad I_{\rm U} = (I_{\rm ord}^{-22.5^\circ} + I_{\rm extra}^{-22.5^\circ} + I_{\rm ord}^{-67.5^\circ} + I_{\rm extra}^{-67.5^\circ})/2$$

are the total intensities in the images used for the calculation of $p_{\rm q}$ and $p_U$. Finally, the radial and tangential polarization are calculated as (also described above):

$$P_\perp=+Q\,{\rm cos}\,2\phi+U\,{\rm sin}\,2\phi\quad;\quad\quad P_\parallel=-Q\,{\rm sin}\,2\phi+U\,{\rm cos}\,2\phi\quad;\quad\quad \phi ={\rm arctan} \frac{x-x_0}{y-y_0}+\theta$$

This is done for each quadruplet of images (four images for four HWP position angles). All individual Stokes $Q$, $U$ $P_\perp$, $P_\parallel$ and $I$ measurements are then mean-combined (ignoring those pixels that have been identified as being in the non-linear detector regime) to yield the final results.

\end{appendix}

\end{document}